# Intrinsic magnetic topological insulators

Pinyuan Wang[1], Jun Ge[1], Jiaheng Li[2,3], Yanzhao Liu[1], Yong Xu[2,3,4], and Jian Wang*[1,2,5,6]

[1]International Center for Quantum Materials, School of Physics, Peking University, Beijing 100871, China

[2]State Key Laboratory of Low Dimensional Quantum Physics, Department of Physics, Tsinghua University, Beijing 100084, China

[3]Frontier Science Center for Quantum Information, Beijing 100084, China

[4]RIKEN Center for Emergent Matter Science (CEMS), Wako, Saitama 351-0198, Japan

[5]CAS Center for Excellence in Topological Quantum Computation, University of Chinese Academy of Sciences, Beijing 100190, China

[6]Beijing Academy of Quantum Information Sciences, Beijing 100193, China

*Correspondence: jianwangphysics@pku.edu.cn

**ABSTRACT**

Introducing magnetism into topological insulators breaks time-reversal symmetry, and the magnetic exchange interaction can open a gap in the otherwise gapless topological surface states. This allows various novel topological quantum states to be generated, including the quantum anomalous Hall effect (QAHE) and axion insulator states. Magnetic doping and magnetic proximity are viewed as being useful means of exploring the interaction between topology and magnetism. However, the inhomogeneity of magnetic doping leads to complicated magnetic ordering and small exchange gaps, and consequently the observed QAHE appears only at ultralow temperatures. Therefore, intrinsic magnetic topological insulators are highly desired for increasing the QAHE working temperature and for investigating topological quantum phenomena further. The realization and characterization of such systems are essential for both fundamental physics and potential technical revolutions. This review summarizes recent research progress in intrinsic magnetic topological insulators, focusing mainly on the antiferromagnetic topological insulator $MnBi_2Te_4$ and its family of materials.

## I. INTRODUCTION

Before the concept of topology was introduced into condensed matter physics, symmetry and symmetry breaking were believed to govern the phase transitions of condensed matter [1]. The discovery of the integer quantum Hall effect (IQHE) [2] paved a new way for classifying matter and understanding phase transitions, and since then various topological systems have been recognized in condensed matter physics [1–17]. In 1980, von Klitzing *et al.* [2] discovered that when a two-dimensional (2D) electron gas is subjected to a strong magnetic field, the longitudinal resistance becomes zero while the Hall resistance shows a quantized plateau with a height of $h/ve^2$, where $h$ is Planck's constant, $v$ is the filling factor, and $e$ is the electron charge. Later, it was revealed that the filling factor $v$ is actually a topological invariant, which is an integral invariant defined at the Brillouin zone and is insensitive to the geometry of the system [2]. This topological invariant is the well-known Thouless–Kohmoto–Nightingale–Nijs invariant [3], also known as the Chern number, which is expressed by the integral of the Berry curvature in the 2D Brillouin zone [1], i.e., $C = \frac{1}{2\pi}\sum_m \int d^2k\, F_m$, where the Berry curvature $F_m = \nabla_k \times A_m$ is the curl of the Berry connection of the $m$th occupied band $A_m$. In 2D systems, the Chern number can be used to classify whether the system is topologically trivial ($C = 0$) or nontrivial ($C \neq 0$). For ordinary insulators, their energy bands are topologically equivalent to those of a vacuum, and so these systems share the same topological invariants as a vacuum and are topologically trivial. However, there exist topologically nontrivial systems that are topologically inequivalent to ordinary insulators. In 2D systems with broken time-reversal symmetry and $C \neq 0$, gapless edge states emerge at the boundary between topologically nontrivial materials and ordinary insulators (vacuum) because of the distinct band topology [1]. The electrons in these edge states can transport only unidirectionally along the boundary and are called chiral edge states [1], wherein elastic backscattering by impurities is forbidden. Therefore, the topological edge states can transport with little dissipation [1,15], thereby offering potential for generating new devices or concepts for low-power-consumption electronics and spintronics [4,5].

Symmetry plays a key role in materials that have topologically nontrivial properties, i.e., topological materials, and the topology of certain systems is protected by certain symmetry, such as time-reversal symmetry [6,17], lattice symmetry [18], and electron–hole symmetry [15,16]. For materials with time-reversal symmetry, the Chern number must be zero [6,7]. In 2005, Kane and Mele [6,7] proposed classifying time-reversal invariant systems by another type of topological invariant known as the $Z_2$ invariant. Here, the terminology "$Z_2$" comes from the fact that this invariant can take only two values, i.e., 0 or 1, depending on the parity of the intersection numbers between the edge states and the Fermi level. Therefore, 2D time-reversal invariant insulators can be divided into two categories: an insulator with $Z_2$ invariant equal to 0 is considered to be topologically trivial, in which the edge states intersect with the Fermi level an even number of times; by contrast, an insulator with $Z_2$ invariant equal to 1 is identified as a topological insulator (TI)[1,6,7], in which the edge states intersect with the Fermi level an odd number of times and always exist in the bulk gap. Currently, it is claimed that HgTe/CdTe quantum wells [9,10], InAs/GaSb quantum wells [19], and some single-layer transition-metal dichalcogenides [20–22] are 2D TIs. Known as helical states, the edge states of a 2D TI comprise a pair of counter-propagating chiral states with spin-momentum locking, wherein electrons that transport in opposite directions carry opposite spins; this feature comes from time-reversal symmetry, which forbids electron backscattering. When the bulk state is an insulating one, the transport signal is contributed mainly by the helical states and gives rise to the so-called quantum spin Hall effect (QSHE) [6–10]. The transport evidence of QSHE in 2D TI is the quantized longitudinal resistance, which has been observed in HgTe/CdTe quantum wells [10], InAs/GaSb quantum wells [19], and monolayer WTe$_2$ [22]. However, the longitudinal-resistance plateau in 2D TIs is imperfect, and some observations are inconsistent with theoretical predictions [23,24]. The $Z_2$ classification can also be extended to three-dimensional (3D) systems, with four $Z_2$ invariants being used to distinguish ordinary insulators from "weak" and "strong" 3D TIs [11]. A weak 3D TI can be regarded as being stacked 2D TIs whose gapless surface states are easily destroyed. A strong 3D TI is more robust, with gapless surface states occurring at all the surfaces. These surface states exhibit a helical spin texture wherein the spin is locked to the momentum [1]. Backscattering for 2D surface states is only forbidden for 180° but not for slightly different angles. At present, the most representative strong TIs are materials in the $Bi_2Te_3$ family [13], whose large bulk energy gap and gapless surface states have been confirmed by angle-resolved photoelectron spectroscopy (ARPES) [12,14].

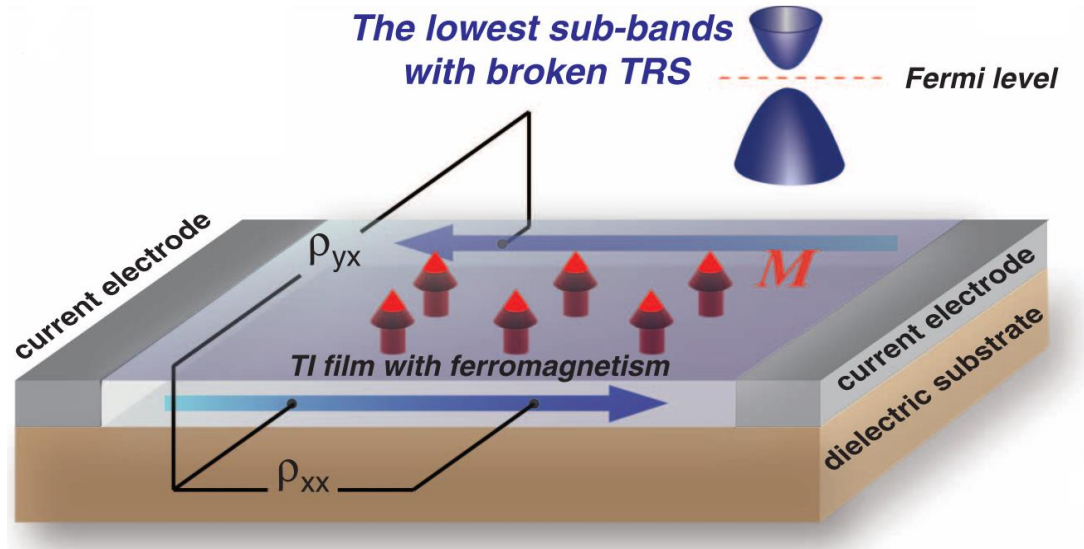

**Figure 1**. Schematic of quantum anomalous Hall effect (QAHE) in a magnetic topological insulator (TI). Illustrated are the exchange gap in the topological surface states of the TI induced by magnetism (denoted by the red arrows) and the resultant chiral edge state. When the Fermi level is tuned to the exchange gap, electronic transport is dominated by the edge state, which results in the QAHE (TRS, time-reversal symmetry). Adapted from [25].

Introducing magnetism into a TI thin film breaks the time-reversal symmetry, and the magnetic exchange interaction can open a gap in the otherwise gapless surface states (Figure 1(a)). Accordingly, the film exhibits a chiral edge state carrying a quantized Hall conductance with a value of $e^2/h$, and this is known as the quantum anomalous Hall effect (QAHE) [25]. Unlike the IQHE, the QAHE requires no external magnetic field and so offers great potential for low-power-consumption electronics.

Magnetic doping [25,26] and magnetic proximity [17] are expected to be useful ways to introduce magnetism into TIs and realize the QAHE. Chang *et al.* [25] were the first to realize the QAHE experimentally in a Cr-doped $(Bi,Sb)_2Te_3$ TI film at ultralow temperatures down to 30 mK. However, in a magnetically doped TI film, the random distribution of magnetic impurities leads to many regions without ferromagnetic (FM) ordering even below the Curie temperature [27]; the electrons in the edge states may scatter with the surface states and the bulk band of the non-FM regions [27], resulting in the overall transport properties of the film deviating from quantization, and so the QAHE can be observed only at ultralow temperatures [25,28]. A method that offers to solve this problem is using a magnetic proximity structure [17], in which the TI film is sandwiched between two FM insulators. However, despite much exploration and effort in that direction, the QAHE is yet to be detected experimentally using that approach. The coupling between the magnetic and topological states is usually weak and depends sensitively on the interface properties, making experimental realization of the QAHE challenging.

As a new and efficient means of incorporating magnetism into TIs, intrinsic magnetic TIs offer a large magnetic exchange gap and homogeneous surface magnetic ordering [29–30], which are highly desirous for investigating the QAHE further and potential applications in low-power-consumption electronics. Also, intrinsic magnetic TIs provide an ideal platform for studying novel topological states such as axion insulator states [17]. Even though magnetically doped heterostructures have been used to construct axion insulator states [31], convincing evidence for axion states is still needed, and the extremely low transition temperature and the complicated preparation of heterostructures might not meet the requirements for further exploration. In intrinsic magnetic TIs, although magnetism can break the time-reversal symmetry, some systems with a certain magnetic ordering can retain an "equivalent" time-reversal symmetry that combines time-reversal operation and other symmetry operations such as $\mathcal{T}C_n$ and $\mathcal{T}\tau$, where $\mathcal{T}$ is the time-reversal operation, $C_n$ is the rotation operation, and $\tau$ is the fractional lattice translation operation. This equivalent time-reversal symmetry provides additional constraints on the bulk band and protects the $Z_2$ topological properties of magnetic TIs. An example is that the $\mathcal{T}\tau$ symmetry operation in the antiferromagnetic (AFM) TI $MnBi_2Te_4$ (MBT) necessarily quantizes the bulk "axion angle" to π, which is very important for forming an axion insulator [29,32].

However, despite AFM TIs having been proposed in 2010 [33], there were no experimentally realizable candidate materials until the proposal of the AFM TI phase in MBT [29,32,34–36]. Rienks *et al.* [30] reported that Mn doping in $Bi_2Te_3$ films results in the spontaneous formation of stoichiometric MBT septuple layers (SLs) rather than the chemical substitution of Bi by Mn [37,38]. These MBT SLs interpolate between $Bi_2Te_3$ quintuple layers to form natural $Bi_2Te_3$/MBT heterostructures. Excitingly, a large topological gap of around 90 meV was observed in this heterostructure, indicating the potential for intrinsic magnetic TIs to realize the QAHE at higher temperatures [28,30]. The AFM TI phase of single-crystal MBT was soon identified experimentally [36]. MBT is a layered material with the $R\bar{3}m$ space group. Monolayer MBT comprises a Te–Bi–Te–Mn–Te–Bi–Te SL, which can be viewed as intercalating an MnTe bilayer into a $Bi_2Te_3$ quintuple layer [29,35,39] (Figure 2(a)). Theoretical calculations show that the ground state of MBT holds FM ordering within the SL and AFM ordering between neighboring SLs [32,34,36]. In the ground state, the magnetic moments of the Mn atoms are aligned with an out-of-plane easy axis ($z$-axis) and are opposite between adjacent layers, which is labeled as the AFM-$z$ state. The magnetic properties are provided mainly by the $3d$ states of Mn, while the topologically nontrivial properties are dominated by the $p$ states of Bi and Te[30]. The long-range FM intralayer coupling, AFM interlayer coupling, and magnetic anisotropy with an out-of-plane easy axis have been confirmed by measurements using inelastic neutron scattering [40-42]. Band-structure calculations have shown that various exotic topological phases may emerge in MBT, including quantum anomalous Hall insulators (QAHIs), axion insulators, and QSHE insulators in thin films, as well as 3D QAHIs, Weyl semi-metals (WSMs), Dirac semi-metals (DSMs), and AFM TIs in bulk, as shown in Figure 2(b)[29]. These 2D or 3D topological states could be manipulated by spatial dimensions and magnetic configurations. Also, the characteristics of van der Waals materials mean that MBT is easily exfoliated into thin flakes, which is convenient for gate modulation and fabrication of various heterostructures. Therefore, intrinsic magnetic TI MBT is an ideal system for investigating emergent rich topological phenomena.

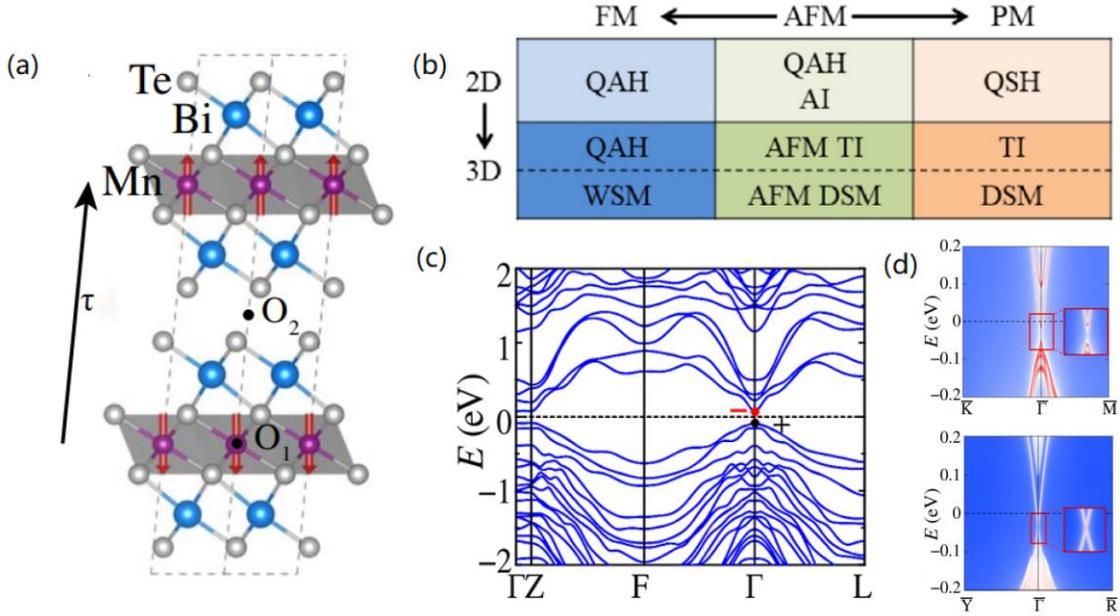

**Figure 2**. Crystal structures and calculated topological properties of $MnBi_2Te_4$ (MBT): (a) the directions of the magnetic moments in each layer are represented by arrows; τ denotes the half-lattice translation operation; (b) various magnetic configurations (FM, ferromagnetism; AFM, antiferromagnetism; PM, paramagnetism) and corresponding topological states of thin-film and bulk MBT (WSM, Weyl semi-metal; DSM, Dirac semi-metal; AI, axion insulator); (c) calculated band structures of bulk AFM MBT; (d) the top surface states open an exchange energy gap induced by magnetic moments under the AFM-$z$ configuration, while all side surface states remain

gapless due to equivalent time-reversal symmetry. Adapted from [29].

## II. AFM AND FM PHASES OF MBT

When exfoliated into thin films, the upper and lower surfaces of few-SL MBT have either the same or opposite magnetic moments according to the parity of the number of layers, thereby determining the topological properties of the system. Also, given that the interlayer AFM interaction in MBT is weak, the interlayer magnetic sequence of MBT can be modulated into out-of-plane FM by applying a moderate out-of-plane magnetic field, whereby MBT becomes the simplest magnetic WSM with only one pair of Weyl points [29,32,43]. In the following, the electronic properties of the MBT system are discussed by considering the number of layers, tunable magnetic configurations, and other parameters.

### A. AFM phase

In the ground magnetic configuration (AFM-$z$), the equivalent time-reversal symmetry $S = \mathcal{T}\tau$ is maintained, where $\tau$ denotes the half-magnetic-lattice translation along the $z$-axis connecting nearest-neighbor SLs with opposite magnetic moments. The existence of the equivalent time-reversal symmetry means that AFM-$z$ MBT can be classified into the $Z_2$ topology. The calculated band structures and related topological invariants show that bulk AFM-$z$ MBT is an AFM TI [29,32,34]. The bulk band opens a topologically nontrivial energy gap after considering spin–orbit coupling (Figure 2(c)). Despite the side surface states remaining gapless, the top surface states open an exchange energy gap below the Néel temperature (~25 K) because of the out-of-plane net magnetic moments (Figure 2(d)) [36].

Few-SL MBT retains the characteristics of interlayer AFM ordering (Figure 3). Therefore, the overall magnetic properties and the corresponding topological properties of few-SL MBT change with the number of layers. Table I gives the calculated thickness-dependent magnetic anisotropy energies (MAEs) and the energy difference between the AFM and FM phases ($\Delta E_{A/F} = E_{AFM} - E_{FM}$) in MBT. All the MAEs are positive, indicating an out-of-plane easy axis in good agreement with experiments. The positive $\Delta E_{A/F}$ of monolayer MBT indicates that the FM phase is the low-energy state, and thus monolayer MBT is a 2D FM material [44]. Furthermore, the calculated band structures indicate that monolayer MBT is topologically trivial (Figure 3(a)). Even-layer MBT is a fully compensated antiferromagnet (cAFM), in which the magnetic moments of each layer are completely compensated because of the AFM nature. However, odd-layer MBT with three or more layers is an uncompensated antiferromagnet (uAFM) exhibiting net magnetization. Table II gives the topological phases and band gaps of monolayer and multi-layer MBT. Odd-layer MBT thin films have Chern insulator states of |$C$| = 1 (Figure 3(c)). The Chern number of even-layer MBT is zero, but it is topologically nontrivial. The upper and lower surfaces of even-layer MBT contribute half quantum Hall conductance with opposite signs, giving zero Hall conductance. This configuration is known as the zero plateau quantum anomalous Hall (ZPQAH) effect state or axion insulator state [32] (Figure 3(b)(d)).

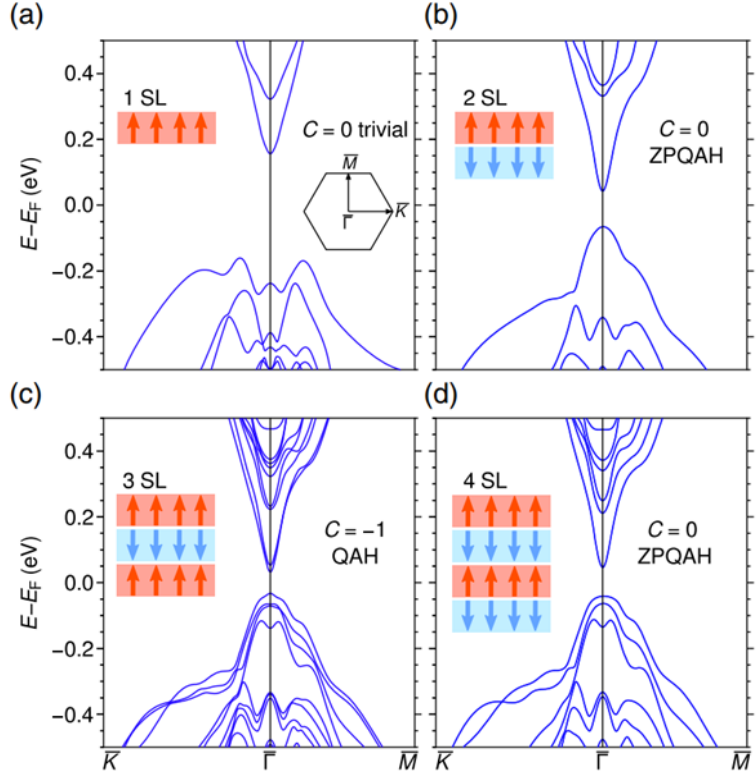

**Figure 3**. Calculated band structures of few-SL AFM MBT: (a) single-layer MBT is topologically trivial, whereas (b) two or (d) more even layers of MBT open a topologically nontrivial gap of around 100 meV that exhibits ZPQAH, and (c) three or more odd layers of MBT show QAHE. Adapted from [34].

**TABLE I.** Thickness dependence of magnetism of MBT films [34]. $T_c$ represents the Curie temperature for monolayer MBT and the Néel temperature for multilayer MBT. The numbers in brackets indicate the error bar.

| Thickness (no. of SLs) | $\Delta E_{A/F}$ [meV/(Mn pair)] | Ordering | MAE [meV/Mn] | $T_c$ [K] |
|---|---|---|---|---|
| 1 | 14.77 | FM | 0.125 | 12(1) |
| 2 | −1.22 | cAFM | 0.236 | 24.4(1) |
| 3 | −1.63 | uAFM | 0.215 | |
| 4 | −1.92 | cAFM | 0.210 | |
| 5 | −2.00 | uAFM | 0.205 | |
| 6 | −2.05 | cAFM | | |
| 7 | −2.09 | uAFM | | |
| ∝ (bulk) | −2.80 | cAFM | 0.225 | 25.42(1) |

**TABLE II.** Thickness dependence of the MBT films' topology and band gap size [34].

| Thickness (no. of SLs) | Topology | Band gap [meV] |
|---|---|---|
| 1 | Trivial | 321 |
| 2 | ZPQAH | 107 |
| 3 | QAH | 66 |
| 4 | ZPQAH | 97 |
| 5 | QAH | 77 |
| 6 | ZPQAH | 87 |
| 7 | QAH | 85 |
| ∝ (bulk) | 3D AFM TI | 225 |

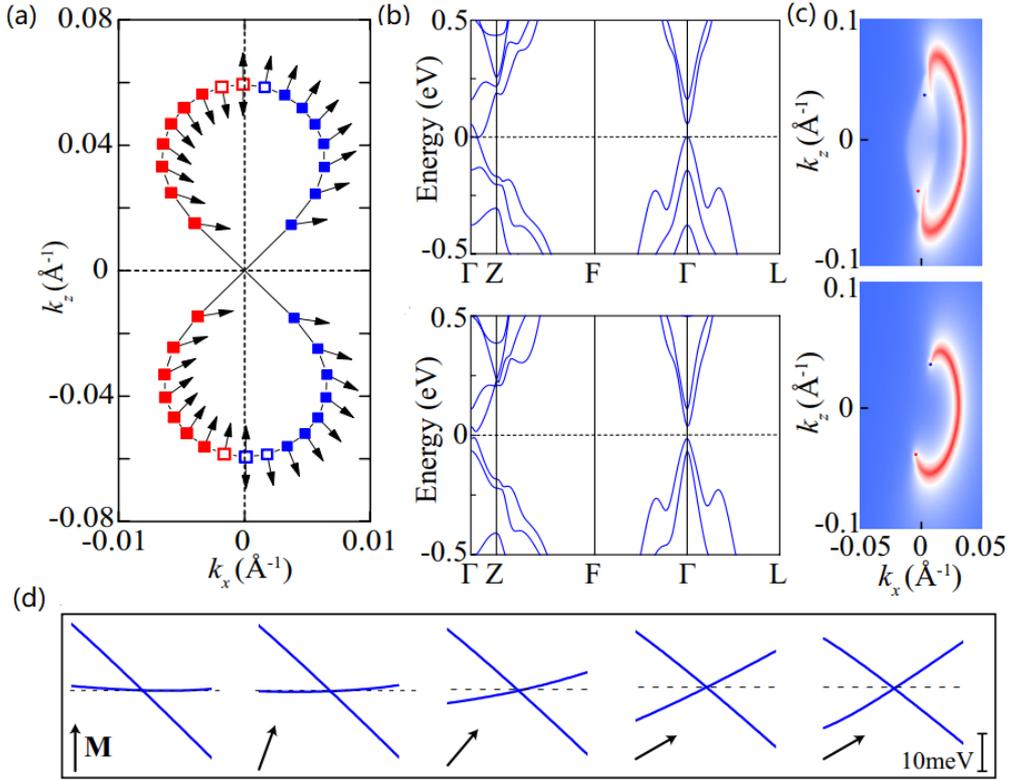

**Figure 4**. Topological phase transition of bulk FM MBT under external magnetic field: (a) phase transition between type-I WSMs (solid squares) and type-II WSMs (open squares); Weyl points evolve in the $k_x$–$k_z$ plane as the direction of the magnetic field (black arrows) rotates from the z-axis to the x-axis; calculated band structures of (b) bulk and (c) surface states of FM MBT with magnetic orientation angles of 10° (upper) and 50° (lower); (d) band dispersion around Weyl points in out-of-plane direction under distinct magnetic configurations with polar angle $\theta$ = 0°, 20°, 40°, 60°, and 80°; the Fermi level is denoted by the dashed line, and the tilted Weyl cone becomes upright gradually with increasing $\theta$. Adapted from [43].

**TABLE III.** Distinct symmetries possessed by bulk MBT with different magnetic configurations [43].

|       | $\mathcal{P}$ | $\mathcal{PT}$ | $M_x$ | $C_{3z}$ | $S = \mathcal{T}\tau$, |
|-------|---|---|---|---|---|
| AFM-$z$ | ✓ | ✓ | ✗ | ✓ | ✓ |
| AFM-$x$ | ✓ | ✓ | ✓ | ✗ | ✓ |
| FM-$z$  | ✓ | ✗ | ✗ | ✓ | ✗ |
| FM-$x$  | ✓ | ✗ | ✓ | ✗ | ✗ |

**B. FM phase**

Under a moderate magnetic field, AFM MBT can be transformed into FM ordering, and accordingly the topological properties also change. The FM-$z$ phase of the bulk material is predicted to be the type-II WSM phase. As given in Table III, unlike the time-reversal invariant WSMs with even pairs of Weyl points, this magnetic WSM breaks time-reversal or equivalent time-reversal symmetry and hosts only a pair of Weyl points. Therefore, FM-$z$ MBT is claimed to be the simplest WSM, which is advantageous for future experimental studies of Weyl physics. Nontrivial transport phenomena including negative longitudinal magnetoresistance [45], large intrinsic anomalous Hall effect [46], and large anomalous Nernst effect [47] can be expected in this WSM phase. The pair of Weyl points is located on the Γ–Z axis, which is protected by threefold rotational symmetry $C_{3z}$. When the external magnetic field is rotated, the bulk system changes continuously from the FM-$z$ phase to the FM-$x$ phase (shown in Figure 4), and the pair of Weyl points deviates from the Γ–Z axis to become general $k$ points in the Brillouin zone because of the broken $C_{3z}$ symmetry. When the polar angle $\theta$ of the external magnetic field is such that $10° < \theta < 20°$, the system changes from type-II WSM to type-I WSM (Figure 4(a)). When $\theta = 90°$, the Weyl points meet and annihilate each other, making the FM-$x$ phase into a trivial FM insulator. Figures 4(b) and 4(c) show the calculated bulk band structures and Fermi arc of MBT with $\theta = 10°$ (upper) and 50° (lower), which are typical of type-II and type-I Weyl points, respectively. The band dispersions of the Weyl points in the out-of-plane direction are shown in Figure 4(d), which also clearly demonstrates the evolution from type-II to type-I WSMs with different magnetic orientations.

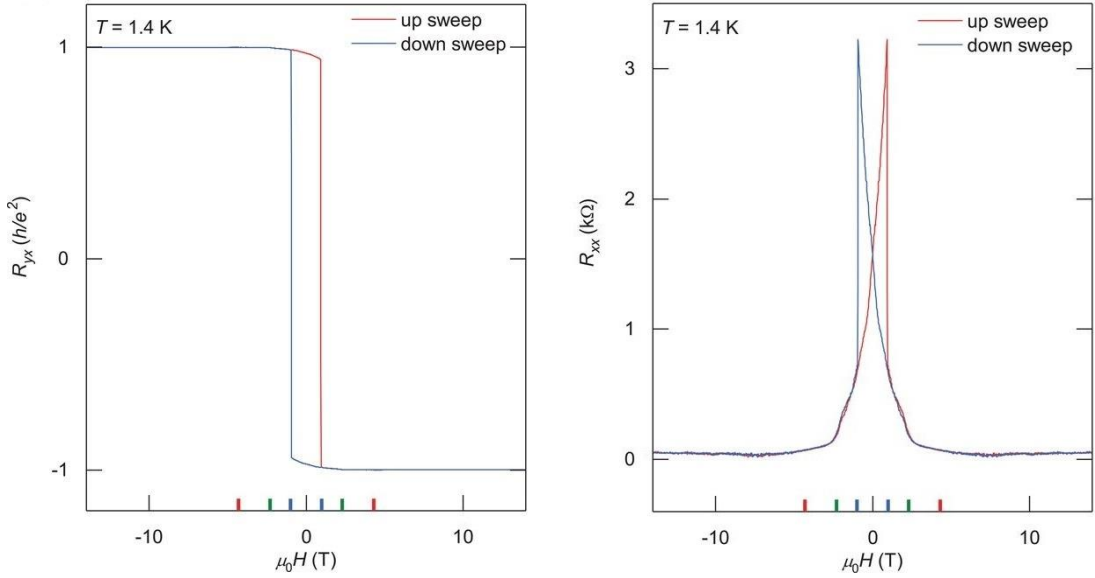

**Figure 5**. QAHE at zero magnetic field in a five-SL MBT flake. $R_{yx}$ and $R_{xx}$ versus magnetic field as measured at 1.4 K. A nearly quantized Hall resistance of $R_{yx} = 0.97 \frac{h}{e^2}$ is observed, and $R_{xx}$ reaches $0.061 \frac{h}{e^2}$. An even better Hall plateau of $0.998 \frac{h}{e^2}$ is obtained as the magnetic field is increased up to 2.5 T. Adapted from [48].

## III. EXPERIMENTAL OBSERVATION OF CHERN INSULATOR WITH HIGH WORKING TEMPERATURE AND HIGH CHERN NUMBER

In previous studies on Chern insulator states (or the QAHE) in magnetically doped TI films, it was necessary to measure at ultralow temperatures because the working temperature of the QAHE is very low [25]. Moreover, only $C = 1$ (one chiral edge state) can be realized in magnetically doped TI films. Therefore, the key issue in the QAHE field is searching for Chern insulators with higher working temperature and high Chern number ($C > 1$), motivated by both fundamental research interest and potential applications in low-power-consumption electronics and spintronics physics.

In a five-SL MBT thin flake, Deng *et al.* [49] reported the nearly quantized Hall resistance of $R_{yx} = 0.97 \frac{h}{e^2}$ at 1.6 K under zero magnetic field, which reached a better plateau of $0.998 \frac{h}{e^2}$ upon increasing the perpendicular magnetic field to 2.5 T (Figure 5). Adopting the quantization criterion of $R_{yx} \sim 0.97 \frac{h}{e^2}$, a quantization temperature of 6.5 K was obtained under an external magnetic field of 7.6 T, which was higher than the previous record of around 2 K by selective doping in TI.

Ge *et al.* [50] reported $C = 1$ Chern insulator states with much higher working temperatures [Figs. 6(a)–6(d)]. In a seven-SL MBT device, Ge *et al.* observed a well-quantized Hall plateau of $R_{yx} = 0.98\frac{h}{e^2}$ at 1.9 K by applying a small back gate of 6.5 V, along with a nearly vanishing resistance of $R_{xx} = 0.012\frac{h}{e^2}$, which is a hallmark of the $C$ = 1 Chern insulator state (Figure. 6(a)(b)). Surprisingly, the quantized Hall plateau shrank slowly with increasing temperature and survived at temperatures as high as 45 K ($R_{yx} \sim 0.904\frac{h}{e^2}$, as shown in Figure. 6(a)), which is the highest working temperature to date for Chern insulator states or the QAHE. Ge *et al.* observed a similar high-working-temperature Chern insulator state in an eight-SL MBT device, where the Hall resistance quantization plateau remained around $0.97\frac{h}{e^2}$ above 30 K and reached $0.997\frac{h}{e^2}$ at 1.9 (Figure. 6(c)(d)). In both samples, the working temperatures of the Chern insulator states were higher than the Néel temperature, indicating that the interlayer AFM coupling is irrelevant for the topological properties of FM MBT and the FM ordering is key for quantizing the Hall resistance. A moderate out-of-plane magnetic field causes the quantization to occur above the Néel temperature by aligning the magnetic moments, and making the FM ordering more robust would allow the QAHE to be realized at temperatures above that of liquid nitrogen.

More interestingly, Ge *et al.* [50] detected a Hall resistance plateau of $0.99\frac{h}{2e^2}$ with a vanishing $R_{xx} \sim 0.004\frac{h}{2e^2}$ in 10-SL MBT devices at 2 K and −15 T by applying a back gate voltage of −58 V ≤ $V_{\text{bg}}$ ≤ −10 V (Figure 7(a)), indicating a high-Chern-number Chern insulator state with $C$ = 2. Figures 7(b) and 7(c) show the temperature evolution of the $C$ = 2 Chern insulator states. With increasing temperature, the Hall resistance plateau remained around $0.97\frac{h}{2e^2}$ at 13 K and $0.964\frac{h}{2e^2}$ at 15 K, while $R_{xx}$ remained below $0.026\frac{h}{2e^2}$ at 13 K and $0.032\frac{h}{2e^2}$ at 15 K. Ge *et al.* also detected the $C$ = 2 state in nine-SL MBT devices. Surprisingly, even for high-Chern-number Chern insulator states, the quantization temperature is much higher than liquid-helium temperatures, suggesting a different mechanism from that of magnetically doped TI films and potential applications in low-energy-dissipation electronics. From Landauer–Büttiker theory [51], although the chiral edge current in a QAHI is dissipationless, there is contact resistance between the electrode and the chiral edge channel. In two-terminal devices, this contact resistance would cause a limited longitudinal resistance with a minimum value of $h/(Ce^2)$, thereby constraining the development of low-power-consumption electronics utilizing the ballistic transport of chiral edge states. One way to solve this problem is to seek the QAHE with large Chern number, which could reduce the contact resistance by a factor of 1/*C*. Undoubtedly, the observation of the $C$ = 2 Chern insulator state provides a good starting point for achieving this goal.

A fundamental issue is the physical origin of the observed high-Chern-number Chern insulator states. In the absence of a magnetic field, bulk MBT is an AFM TI in which the side surface states remain gapless while the top and bottom surface states are gapped by magnetic moments. For MBT thin films, the gapped top and bottom surface states carry a half quantum Hall conductance of opposite or identical signs, leading to $C = 0$ or 1. Chern insulator states with $C > 1$ are impossible physically with TI thin films. However, by applying a moderate magnetic field, bulk MBT becomes an FM WSM and thin-film FM MBT exhibits a Chern-insulator band structure when the thickness is within the quantum confined regime, wherein the energy gap decreases and the Chern number may increase as the film gets thicker [43,52]. In other words, as a quantized system, thin-film FM MBT could have high Chern number ($C > 1$) with increasing thickness. This unusual feature can be attributed to the combination of the quantum confinement effect and the WSM characteristics of the MBT FM phase, which paves a new way to realize high-Chern-number Chern insulators (Figure 7(d)) [53,54]. For $N$-SL FM MBT, its electronic states can be viewed as the quantum-well states of an FM WSM and possess a finite band gap due to quantum confinement effects. The anomalous Hall conductance of such a system would be estimated by $\sigma_{xy} \approx \frac{c_0 N k_w e^2}{\pi h}$, where $c_0$ is the out-of-plane thickness of each SL, $N$ is the number of SLs, and $k_w$ is half the distance between the Weyl points at the Brillouin zone (Figure 7(e))[50]. For a gapped 2D topological thin film, $\sigma_{xy}$ must take quantized values as $\sigma_{xy} = \frac{C(N)e^2}{h}$, therefore the Chern number is determined approximately by $C(N) \sim \frac{c_0 N k_w}{\pi}$. Upon increasing the layer number $N$ by $\pi/|c_0 k_w|$ (around four SLs according to first-principles calculations), the Chern number increases by 1 (Figure 7(f)), which is consistent with the experimental observations. The experimental discovery of the $C = 2$ Chern insulator state in MBT devices also offers indirect evidence for the magnetic WSM phase in FM-$z$ MBT.

Topological edge states in the QAHE can maintain the quantum characteristics of electrons (such as no dissipation) at the macroscopic scale and may be used to design and construct electronic devices based on new principles. The high-working-temperature and high-Chern-number Chern insulator states discovered in MBT indicate that if the proper intrinsic magnetic topological materials and external parameters are chosen, then there is hope of realizing the QAHE at temperatures above that of liquid nitrogen and perhaps even at room temperature, which would be a milestone breakthrough in practical applications of the QAHE.

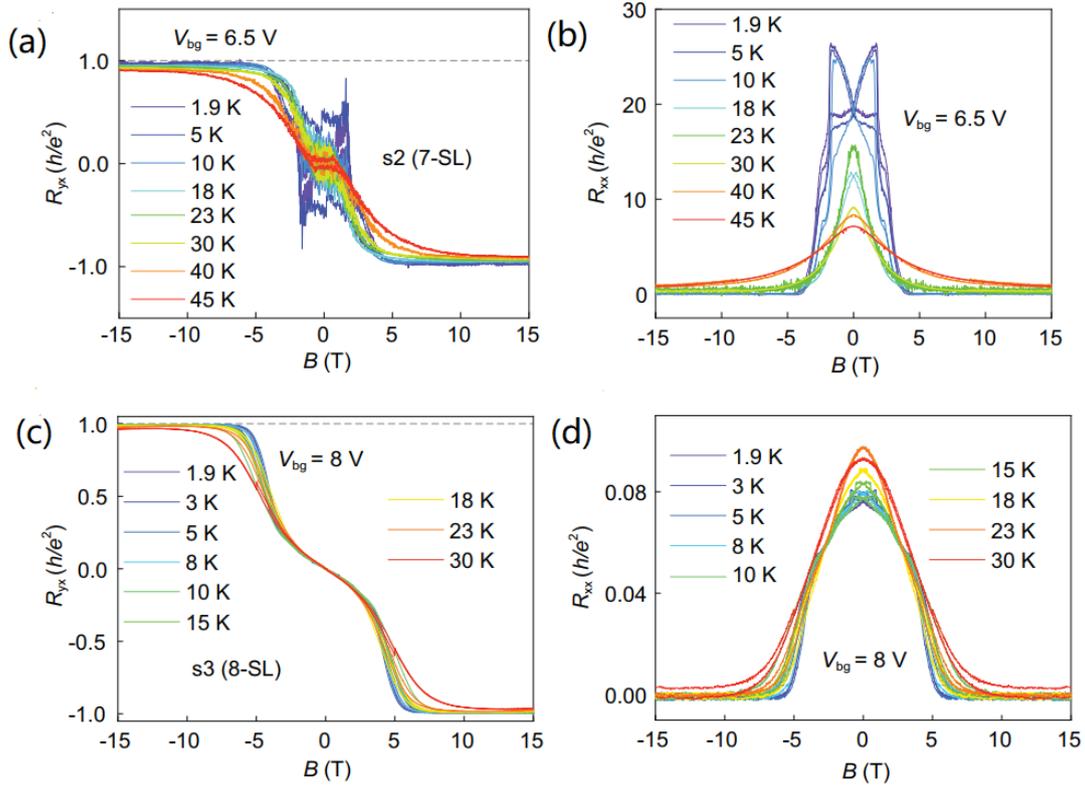

**Figure 6**. High-temperature Chern insulators in (a), (b) seven-SL and (c), (d) eight-SL MBT devices: (a), (b) temperature dependence of $C = 1$ Chern insulator states in a seven-SL MBT device; the nearly quantized Hall resistance plateau remains at temperatures as high as 45 K; (c), (d) temperature dependence of $R_{yx}$ and $R_{xx}$ versus magnetic field in an eight-SL MBT device; a well-defined quantized Hall resistance plateau survives up to 30 K. Adapted from [50].

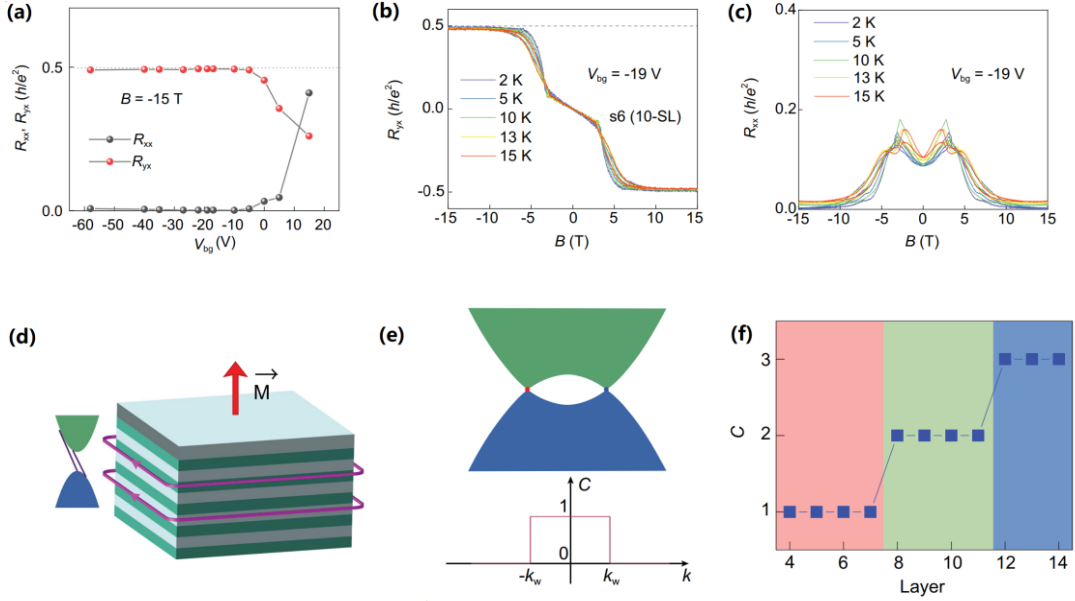

**Figure 7**. High-Chern-number Chern insulator states with *C* = 2 in 10-SL MBT device: (a) $R_{xx}$ and $R_{yx}$ as functions of back gate voltage at 2 K and −15 T, featured by a Hall resistance plateau of $h/2e^2$ and vanishing $R_{xx}$ at back gate voltage of −58 V ≤ $V_{bg}$ ≤ −10 V, indicating a Chern insulator with Chern number *C* = 2; (b), (c) temperature dependence of high-Chern-number Chern insulator states in 10-SL MBT device; $R_{yx}$ and $R_{xx}$ at different temperatures from 2 K to 15 K are shown as functions of the magnetic field strength; a Hall resistance plateau of $0.97h/2e^2$ survives to 13 K; (d) schematic of high-Chern-number Chern insulator states with two chiral edge states across the band gap; gray and green indicate adjacent MBT SLs; (e) illustration of band structure of bulk FM MBT, which is a magnetic WSM; the distance between the Weyl points at the Brillouin zone is $2k_w$, and the Chern number jumps at the positions of the Weyl points; (f) calculated Chern number as a function of film thickness. Adapted from [50].

## IV.    EXPERIMENTAL OBSERVATION OF AXION INSULATOR

In the context of quantum chromodynamics, considering the axion field introduces additional terms in Maxwell's equations that couple the electric and magnetic fields. Interestingly, by using effective topological field theory, it is found that the electromagnetic response of a 3D TI includes the additional term $S = \frac{\theta}{4\pi^2}\alpha \int d^4x\, \boldsymbol{E} \cdot \boldsymbol{B}$ [17], which is similar to Maxwell's equations having additional axion terms. Here, $\boldsymbol{E}$ and $\boldsymbol{B}$ are the electromagnetic fields, $\alpha$ is the fine-structure constant, and the so-called axion angle $\theta$ is a dimensionless pseudoscalar parameter that is defined modulo $2\pi$ [32]. Also known as the axion term, the extra term couples the magnetic and electric fields and leads to the topological magnetoelectric effect (TME) [32]. From the effective action of a 3D TI, the current response term is $j = j_{free} + \frac{\alpha}{4\pi^2}\nabla\theta \times \boldsymbol{E} - \frac{\alpha\theta}{4\pi^2}\frac{\partial \boldsymbol{B}}{\partial t}$. Note that the second term therein is the Hall current (perpendicular to the electric field) and is due entirely to the axion term, which is the result of the nontrivial topology of the system. If the energy gap is fully opened on the surface and the bulk-state axion angle is strictly quantized to π, then the axion term will be the only current contribution to the stable system. Integrating along the surface normal direction, we obtain the half quantum conductance $\sigma_{xy} = e^2/2h$ contributed by each surface, where the sign is determined by the direction of the surface magnetic moments. When the directions of the magnetic moments are opposite between the top and bottom surfaces, the current directions of the axion terms are also opposite (considering the local coordinates of the upper and lower surfaces). Therefore, although a certain surface has half-integer Hall conductance, the net Hall current of the system becomes zero because of the cancellation of the up and down surfaces, and so ZPQAH can be observed [34]. Furthermore, a circulating Hall current is formed when applying an electric field, as shown in Figure 8. Such a circulation is equivalent to the surface magnetizing current induced by a quantized magnetization $\boldsymbol{M} = -\left(n+\frac{1}{2}\right)\frac{e^2}{hc}\boldsymbol{E}$ wherein $n$ is an integer. Similarly, a charge polarization can be induced by applying a magnetic field. When a magnetic field is turned on slowly, the induced electric field will generate a Hall current parallel to the magnetic field and thus charges will accumulate on the top and bottom surfaces, which is equivalent to a quantized polarization $\boldsymbol{P} = \left(n+\frac{1}{2}\right)\frac{e^2}{hc}\boldsymbol{B}$. Such striking induction between electric and magnetic fields is known as the TME, which is direct evidence for axion insulator states. However, the strict requirements of instrument accuracy and sample quality make it quite challenging to observe the TME. Therefore, the research effort to date has been focused on detecting the zero Hall platform as evidence for the axion insulator phase [31,55].

To realize the axion insulator states, three conditions must be satisfied [17,32]: (i) all the topological surface states of the system must be fully gapped; (ii) there must be a certain symmetry to quantize the axion angle $\theta$; (iii) the Fermi surface must fall into the energy gaps of the bulk state and all the surface states. FM/TI/FM heterostructures have been used to construct axion insulators, such as V-doped $(Bi,Sb)_2Te_3$/$(Bi,Sb)_2Te_3$/Cr-doped $(Bi,Sb)_2Te_3$ heterostructures [31]. The principle is to make use of the different coercive fields of the magnetic films on the top and bottom FM sides of the heterostructure, thereby making the magnetic moment directions of the upper and lower surfaces of the system opposite under a specific external magnetic field. However, the working temperature of the system is of the order of millikelvins and preparing the heterojunction is complicated, thereby not favoring widespread investigations.

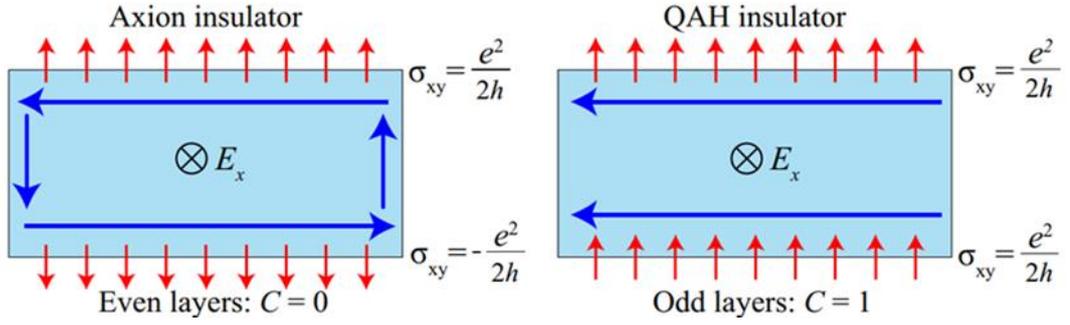

**Figure 8**. Edge states of axion insulator and QAHI. Directions of magnetic moments are denoted by red arrows and the Hall currents are represented by blue arrows. Each gapped surface contributes a half-quantized Hall conductance, the signs of which are opposite in axion insulator states and the same in QAHI states. In axion insulator states, an electric field can induce a quantized circulating Hall current that is equivalent to the surface magnetizing current induced by a quantized magnetization. Adapted from [29].

As an intrinsic magnetic TI, MBT is expected to overcome these difficulties. Its ordered structure makes MBT more advantageous for avoiding disorder effects such as band-gap fluctuation and superparamagnetism [32]. The large band gap facilitates the observation of axion insulator states at higher temperature. The AFM-$z$ phase of even-layer MBT thin films is an intrinsic axion insulator because the top and bottom surface states are naturally gapped by opposite magnetic moments. Liu *et al.* [55] reported a large longitudinal resistivity with a weak magnetic field at 1.6 K by applying a gate voltage of 25 V (Figure 9(a)). More interestingly, the Hall resistivity remained zero in a field range of −3.5 T < $B$ < 3.5 T. Figure 9(b) shows (i) the zero-field longitudinal resistivity and (ii) the slope of the weak-field Hall resistivity versus the magnetic field strength as functions of the gate voltage $V_g$. For $V_g$ = 22–34 V, the longitudinal resistivity exhibits insulating behavior while the slope of the Hall resistivity exhibits a zero plateau, which is different from conventional insulators. With increasing magnetic field, both the longitudinal and Hall resistivities undergo a sharp transition. The Hall resistivity reaches a plateau of $0.984\frac{h}{e^2}$ with a longitudinal resistivity of $0.018\frac{h}{e^2}$ at −9 T (Figure 9(a)). As discussed above, the zero Hall plateau indicates the appearance of the axion insulator phase, while the quantized Hall plateau and vanishing longitudinal resistance are characteristics of Chern insulator states with Chern number $C$ =1. Therefore, a magnetic-field-driven phase transition between axion and Chern insulator states is observed.

Recently, two independent studies reported unusual layer-dependent magnetic properties in MBT devices [56,57], observing anomalous magnetic hysteresis loops in polar reflective magnetic circular dichroism spectroscopy (RMCD) results [56] and anomalous Hall curves in even-layer MBT [57]. More intriguingly, Ovchinnikov *et al.* [57] found that the hysteresis loop near zero magnetic field vanished for an odd-layer device in the AFM state. Actually, most device samples in experiments are inevitably affected by various factors during device fabrication process and measurements, including sample quality and oxidation. Consequently, the properties of devices depend sensitively on the specific experimental conditions and might vary between different samples.

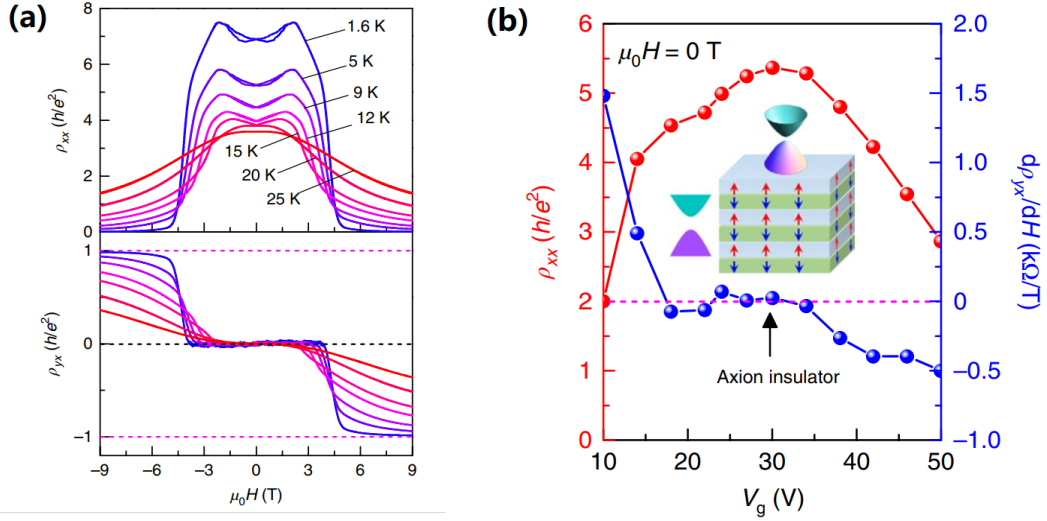

**Figure 9**. Magnetic-field-driven axion-insulator–Chern-insulator transition in a six-SL MBT device: (a) longitudinal and Hall resistivities versus magnetic field strength at various temperatures with gate voltage $V_g = 25$ V; (b) gate dependence of axion insulator state. Adapted from [55].

## V.  MBT-RELATED SYSTEMS

### A.  Heterostructures comprising MBT and magnetic monolayer materials

Constructing a heterostructure from MBT thin flakes and other 2D magnetic materials has been proposed as a promising way to stabilize the surface moments of MBT and realize high-temperature QAHE. Fu *et al.* [58] calculated the structural characteristics of an MBT–CrI$_3$ heterostructure. They found that the CrI$_3$ monolayer had strong FM exchange interaction with the MBT surface layer, effectively increasing the stability of magnetic ordering in the MBT surface layer. The CrI$_3$ monolayer induced an exchange bias as large as 40 meV, which is much larger than the Néel temperature of MBT (around 25 K, which corresponds to an energy of around 2 meV [36]). Moreover, from band-structure calculations, the proximity with the CrI$_3$ monolayer had little effect on the band topology of MBT, as shown in Figure 10(a). Therefore, heterostructures of MBT thin films and CrI$_3$ monolayers could help to realize the QAHE at higher temperatures.

### B.  MBT–superconductor heterostructures

The realization of Majorana fermions in condensed matter systems, including Majorana zero energy modes and Majorana chiral edge states, has been of great concern [4,15,16,59–62].

Majorana zero energy modes can be used to construct quantum bits and have important applications in topological quantum computation [59]. One-dimensional (1D) Majorana chiral edge states are topological edge states of 2D $p+ip$ chiral superconductors. Fu and Kane [60] proposed heterostructures comprising FM magnets, TIs, and s-wave superconductors to form equivalent $p+ip$ superconductors, wherein the possible 1D Majorana chiral edge states can be obtained. Furthermore, Qi *et al.* [16] suggested that a chiral topological superconductor carrying a Majorana chiral state might be promising in a QAHI in proximity to an s-wave superconductor. Based on this proposal, He *et al.* [63] claimed to realize Majorana chiral edge states characterized by an $e^2/2h$ conductance plateau in Cr-doped $(Bi,Sb)_2Te_3$ thin films in proximity with an Nb superconductor, but the results are questionably extrinsic [64–67]. Therefore, it is important to obtain equivalent $p+ip$ superconductors by using intrinsic magnetic topological system. Peng *et al.* [68] proposed a realization of chiral Majorana states using a heterostructure comprising MBT and an s-wave superconductor. In such a heterostructure, the superconducting proximity effect can open a superconducting gap on the MBT side surface. This energy gap and the magnetic exchange energy gap of the upper and lower surfaces are topologically distinct, and the hinge of the heterostructure will exhibit 1D Majorana chiral edge states (as shown in Figure 10(b)).

**C. MBT family**

In addition to MBT, a large family of van der Waals materials that can be expressed as $MnBi_2Te_4(Bi_2Te_3)_n$ ($n$ = 1,2,3 …) also have intrinsic magnetic topological bands [69–80]. In these materials, the MBT layers are separated by $n$ layers of $Bi_2Te_3$, as shown in Figure 10(c). The interlayer AFM interactions decrease quickly as the distance between MBT layers increases, so the magnetic and topological properties are highly tunable by the number $n$ of $Bi_2Te_3$ layers. For larger $n$, the AFM coupling is weaker and FM ground states are favored. When $n$ = 1 or 2, $MnBi_4Te_7$ and $MnBi_6Te_{10}$ retain interlayer AFM coupling with Néel temperatures of 13 K and 11 K, respectively [77,78]. Transport measurements, ARPES experiments, and band calculations have shown that AFM $MnBi_4Te_7$ is an AFM TI below the Néel temperature, with gapless side surface states similar to MBT. However, the top and bottom surface states are associated with the MBT or $Bi_2Te_3$ termination. The energy gaps of $MnBi_4Te_7$ top surface states with MBT termination are much smaller than those with $Bi_2Te_3$ termination [78]. For $n$ = 2, $MnBi_6Te_{10}$ is an AFM TI below the Néel temperature, having the full topological bulk gap with gapless side surface states and gapped top and bottom surface states [69]. More interestingly, the FM state emerges when applying a magnetic field as weak as 0.1 T and is preserved when the magnetic field is lowered to zero at 2 K. The band structures in the FM state show a gap of around 0.15 eV. Theoretical calculations show that bulk FM $MnBi_6Te_{10}$ is a high-order TI with $Z_4$ = 2, equivalent to an axion insulator with $\theta = \pi$ [69]. Moreover, theoretical calculations of few-layer FM $MnBi_6Te_{10}$ show that two, three, and four layers of $MnBi_6Te_{10}$ ($[Bi_2Te_3]$–$[MnBi_2Te_4]$–$[Bi_2Te_3]$–…) are intrinsic Chern insulators with $C$ = 1 at zero magnetic field [69]. For $n$ = 3, experimental transport results show that $MnBi_8Te_{13}$ falls into the FM phase below 10.5 K, and first-principles calculations and ARPES measurements further demonstrated that $MnBi_8Te_{13}$ is an intrinsic FM axion insulator with $Z_4 = 2$ [70]. Interlayer exchange interaction between neighboring MBT SLs becomes too weak when the layer number $n$ of intercalated $Bi_2Te_3$ layers is increased further. Below the critical temperature, the MBT SLs can be regarded as being magnetically independent from each other, and the magnetization of the SLs becomes disordered along the $z$-axis.

The nontrivial topology of the $(MnBi_2Te_4)(Bi_2Te_3)_n$ family is connected strongly with the corresponding magnetic properties in this series, thereby allowing us to regulate the strength of the interlayer AFM coupling in particular and so control the topological states. For large $n$, small or even zero external magnetic field is required to achieve Chern insulator states because of the weaker interlayer AFM coupling. This fact could also help to realize a topological superconductor hosting Majorana fermions based on the MBT family [68].

In addition to the $(MnBi_2Te_4)(Bi_2Te_3)_n$ family, the intrinsic van der Waals material $Mn_2Bi_2Te_5$—as a sister compound of MBT—has also been predicted to be a topological AFM material, one that could host the long-sought dynamical axion field [81]. Unlike AFM MBT, in which a certain symmetry fixes the axion field $\theta$ to a quantized value ($\theta = \pi$), in $Mn_2Bi_2Te_5$ $\theta$ becomes a bulk dynamical field from magnetic fluctuations and varies from zero to $\pi$ because of the breaking of both time-reversal and inversion symmetries [81]. Such a dynamical axion field could induce novel phenomena, such as axionic polaritons [82] and nonlinear electromagnetic effects induced by axion instability [83]. A similar dynamical axion insulator state was also predicted in an $(MnBi_2Te_4)_2(Bi_2Te_3)$ heterostructure [84]. The dynamical axion insulator states realized in these MBT-related systems may pave the way to a new generation of axion-based devices. The MBT family thus provides a feasible and highly tunable 2D platform for exploring novel quantum topological phases [48,50,55], low-power-consumption electronics, spintronics [4.5], and topological quantum computing [59].

### D. Chemically substituted MBT

Another route to regulating the properties of MBT is element substitution [52,85-87]. Both magnetic and carrier properties are tunable by substituting Sb for Bi, and thus topological effects can be controlled by changing the proportion of Sb atoms in $Mn(Sb_xBi_{1-x})_2Te_4$ [85,86]. Yan *et al.* [85] reported that with increasing Sb content in $Mn(Sb_xBi_{1-x})_2Te_4$, the Néel temperature decreased slightly from 24 K for $MnBi_2Te_4$ to 19 K for $MnSb_2Te_4$, while the critical magnetic field strengths required for spin-flop transition and moment saturation decreased dramatically. The above results suggest that both the interlayer AFM coupling and magnetic anisotropy weaken with increasing Sb content for $x < 0.86$. Besides, Sb substitution can also control the Fermi level. With increasing $x$, a crossover from $n$-type to $p$-type conducting behavior is observed in bulk $Mn(Sb_xBi_{1-x})_2Te_4$ around $x \approx 0.63$ [85]. A critical issue affecting the experimental observation of the high-temperature QAHE in MBT is the bulk $n$-type conductivity, usually induced by high densities of Mn and Bi antisite defects [88]. Sb substitution could modulate the carrier concentration and tune the Fermi level to the bulk gap [85]. Another strategy to lower the Fermi level is to grow higher-quality MBT single crystals under Te-rich conditions [88].

Theoretical calculations predict that MnSb$_2$Te$_4$ is a topologically trivial AFM material [32,86], but experiments report both AFM ($T_N$~19 K) [85] and FM ($T_C$~25 K) [89] MnSb$_2$Te$_4$. Furthermore, Wimmer *et al.* [87] conducted detailed experimental explorations combining ARPES, scanning tunneling microscopy (STM), magnetometry, and x-ray magnetic circular dichroism in epitaxial MnSb$_2$Te$_4$. They observed that MnSb$_2$Te$_4$ exhibits robust out-of-plane FM ordering with a Curie temperature of 45–50 K, and through ARPES and STM they uncovered a Dirac point near the Fermi level with out-of-plane spin polarization and a magnetically induced gap of around 17 meV that vanishes at the Curie temperature. These are typical features of an FM TI. By conducting further density-functional theoretical calculations and x-ray diffraction experiments, Wimmer *et al.* concluded that the FM and topologically nontrivial properties of MnSb$_2$Te$_4$ originate from the antisite mixing of Mn and Sb. The fact that the Curie temperature in MnSb$_2$Te$_4$ is much higher than the Néel temperature in MBT may pave a new way to realizing the high-temperature QAHE.

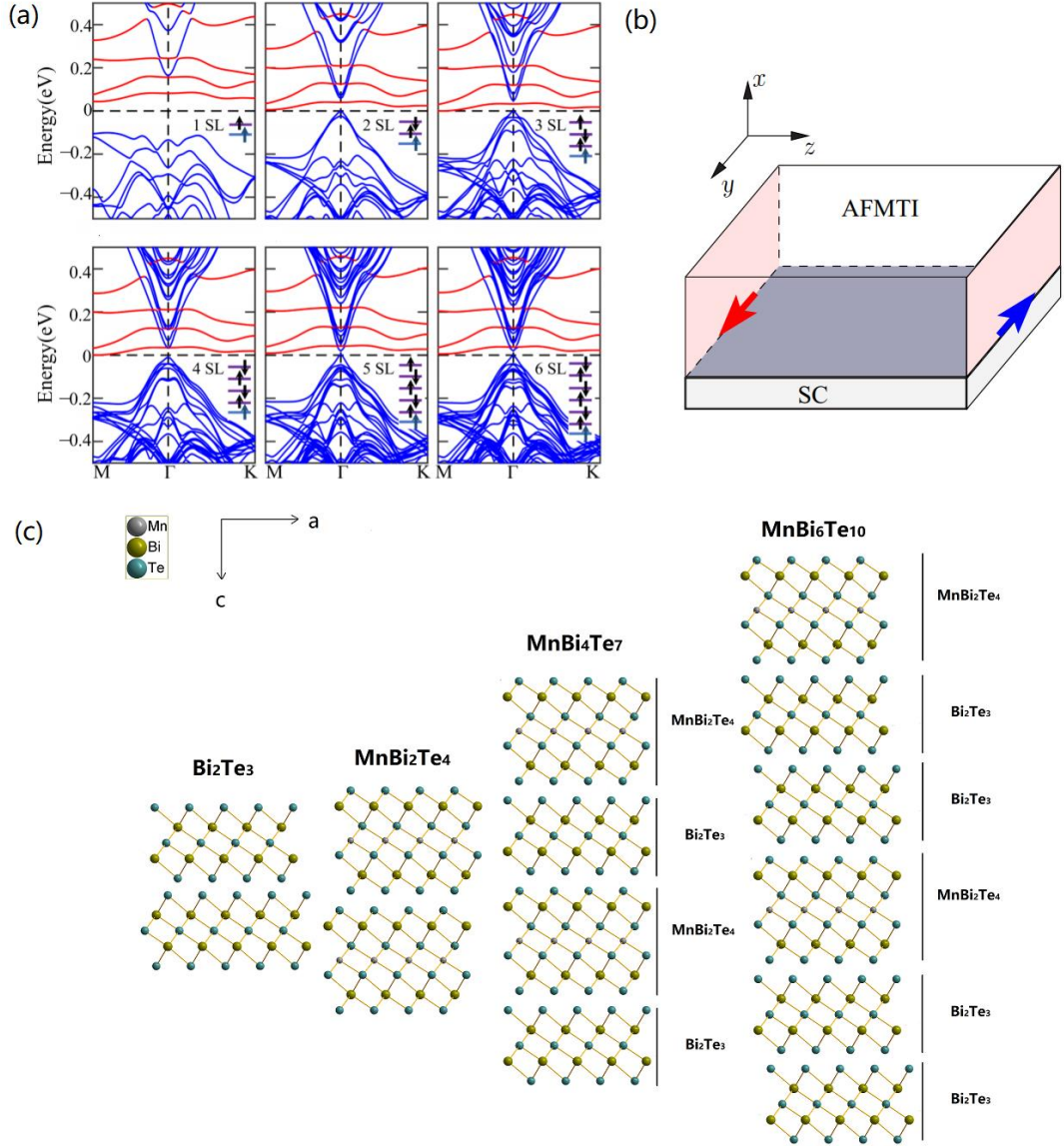

**Figure 10.** MBT heterojunctions and MBT family: (a) calculated band structures of one-SL to six-SL MBT thin films in proximity to a CrI$_3$ layer; the blue lines represent MBT bands and the red lines represent Cr-e$_g$ bands; in the insets, the MBT layers and CrI$_3$ monolayer are denoted by purple and blue layers, respectively; the results show that the CrI$_3$ monolayer induces an exchange bias of around 40 meV, and the band topology of MBT is little affected; adapted from [58]; (b) Majorana hinge modes at the edge of the interface between an AFM TI (e.g., MBT) and an s-wave superconductor; the Majorana modes are indicated by blue and red arrows, and the easy axis of AFM is in the $z$ direction; adapted from [68]; (c) crystal structures of MBT family; the common structural characteristic is that one MBT SL is sandwiched by $n$ Bi$_2$Te$_3$ quintuple layers; shown are the cases of $n = 0$ for MBT, $n = 1$ for MnBi$_4$Te$_7$, and $n = 2$ for MnBi$_6$Te$_{10}$.

## VI. PUZZLE OF SURFACE STATES OF MBT

ARPES experiments have purportedly shown the exchange gap of surface states induced by the out-of-plane magnetic moments below the Néel temperature as having values from tens of millielectronvolts to 200 meV [36,90,91]. However, it is puzzling that the observed surface energy gaps vary little with temperature. Even above the Néel temperature, a limited "energy gap" can still be detected [91], so the energy gap might not originate from the magnetic order. On the contrary, some high-resolution ARPES experiments have reported the top and bottom surfaces as having a gapless surface state [86,92–96] with a small evolution with temperature (Figure 11(a)). The above results are obviously inconsistent with the theoretical predictions for the surface gap opened by the out-of-plane magnetic moments. Li *et al.* [93] attributed the gapless states to the limited interaction between the Mn-3$d$ state that provides the magnetism and the Bi/Te-$p$ orbital state that dominates the topological properties. Using resonance ARPES technology, the Mn-3$d$ state was found to be located at 4 eV below the Fermi energy and is negligible in the energy range in which the nontrivial topology arises [93]. Therefore, Li *et al.* claim that it may be difficult for magnetism to open the exchange energy gap of the surface state. However, this conclusion is yet to be supported by first-principles calculations and contradicts the Chern insulator phase and the axion insulator states observed by transport measurements [48,50,55]. Besides, a large magnetic gap of around 90 meV at the Dirac point in $Bi_2Te_3$/MBT heterostructures has been reported, which vanishes above the Curie temperature [30]. This result, on the other hand, indicates that the interaction between magnetism and topology in MBT may be strong enough to induce an observable magnetic gap, and the absence of the magnetic gap in previous experiments calls for alternative explanations.

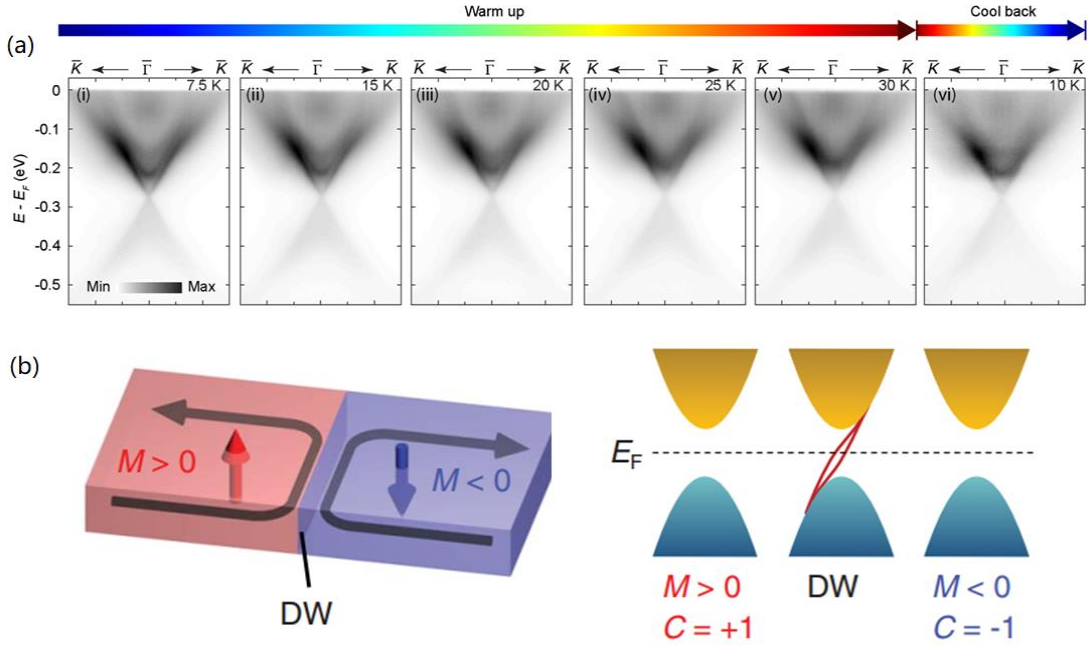

**Figure 11**. (a) Gapless surface states observed by ARPES at various temperatures. Adapted from [92]. (b) Magnetic-moment fluctuation model, which considers that the magnetic-moment fluctuation forms a series of disordered domains: boundary states can be formed between disordered domains with different moments, and these edge states provide the density of electronic states at the Dirac point. Adapted from [97].

Chen *et al.* [92] showed that the observation of gapless surface states can be explained by spatially dependent magnetic moments on the surface. As shown in Figure 11(b), the AFM coupling between the surface layer and the underlying layer may be weaker than the AFM coupling of the bulk state, which leads to fluctuation of the magnetic moment on the surface, forming a series of magnetic-moment domains with the same or opposite directions. The opposite magnetic moments induce gapped surface states with gaps of opposite sign, so gapless edge states crossing the energy gap appear between the domain. These edge states provide the electron density of states for the observation of the gapless surface state in the ARPES experiment. A similar explanation is also proposed [52], in which the directions of surface magnetic moments change in different terraces (domains). Consequently, topological edge states occur at the step edges of different terraces. The aforementioned hypotheses can be justified further by more spatially resolved spectroscopy measurements, such as scanning tunneling spectroscopy and point-contact technology. Recently, a point-contact experiment reported a surface gap of around 50 meV, which vanished as the sample became paramagnetic with increasing temperature [98]. Because point-contact technology studies the local properties of samples, this result indicates the existence of a local magnetic gap at the MBT surface.

## VII. OUTLOOK

The intrinsic magnetic TIs provide a promising platform to explore the interplay between topology and magnetism, which could open up new ways for novel fundamental physics as well as give rise to potential technical revolutions. Theoretical studies have predicted several intrinsic magnetic TIs, such as the MBT family, $EuSn_2As_2$, and $EuIn_2As_2$ [36,93,99]. Among these materials, MBT has attracted the most attention and is believed to have the greatest potential, this being because MBT brings hope for many novel topological quantum states, such as Chern insulator states and axion insulator states [29,32,34,48-50,55]. At moderate magnetic fields, MBT thin flakes translate into FM WSM states with only one pair of Weyl points, in which high-working-temperature and high-Chern-number Chern insulators are realized [50]. Besides, 2D heterostructures based on the MBT family could offer unprecedented opportunities to discover new fundamental physics of topological fermions such as the long-sought Majorana fermions [68]. As a van der Waals magnet, MBT also paves an ideal avenue to exploring novel magnetic phases such as the skyrmion phase [40,44].

Notably, the QAHE with high working temperature can be expected in MBT. The key factor that limits the working temperature of the QAHE in MBT is the low magnetic-ordering temperature, which is restricted by the strong magnetic fluctuation caused by weak magnetic interaction in MBT. Applying a moderate external magnetic field can effectively increase the anisotropy of MBT, suppress the magnetic fluctuation, and thus increase the effective magnetic-ordering temperature [50]. Another way to realize the high-temperature QAHE in MBT is to strengthen the interlayer coupling in MBT by magnetic doping. It is predicted that V-doped MBT is FM with a high Curie temperature of around 45 K [100]. However, magnetic doping may introduce additional magnetic inhomogeneity and thus cause more scattering that may destroy the QAHE. Another way to improve the magnetic properties of MBT is chemical substitution. It is reported that $MnSb_2Te_4$ with slight Mn/Sb antisite mixing is an FM TI with a Curie temperature of 45–50 K [87]. Constructing heterostructures of MBT with magnetic monolayer materials is also expected to be a promising route to suppressing the magnetic fluctuation. Magnetic monolayer materials may effectively increase the stability of the magnetic ordering in the MBT surface layer and thus raise the QAHE working temperature [58]. Once the QAHE is realized above the liquid-nitrogen temperature, a milestone breakthrough for the practical application of topological materials can be expected, which would stimulate research interest significantly in the fields of physics, materials, and information technology.

**Acknowledgements**

This work was financially supported by the Beijing Natural Science Foundation (Grant No. Z180010), the National Key R&D Program of China (2018YFA0305600, 2017YFA0303302), the National Natural Science Foundation of China (Grant No. 11888101, Grant No. 11774008), the Strategic Priority Research Program of Chinese Academy of Sciences (XDB28000000), the Open Research Fund Program of the State Key Laboratory of Low-Dimensional Quantum Physics, Tsinghua University (Grant No. KF202001).

**Competing Interests**

The authors declare no competing interests.